\magnification1200


\vskip 2cm
\centerline
{\bf Supersymmetry anomalies and the Wess-Zumino Model in a supergravity background}
\vskip 1cm
\centerline{ Giorgos Eleftheriou   and Peter West}
\centerline{Department of Mathematics}
\centerline{King's College, London WC2R 2LS, UK}
\vskip 2cm

\vskip1cm
\leftline{\sl Abstract}
We briefly recall the procedure for computing the Ward Identities in the presence of  a regulator which violates the symmetry being considered. We compute the first non-trivial correction to the supersymmetry Ward identity of the Wess-Zumino model in  the presence of background supergravity using dimensional regularisation. We find that the result can be removed using a finite local counter term and so there is no supersymmetry anomaly. 
\vskip2cm
\noindent

\vskip .5cm

\vfill
\eject

Supersymmetry was discovered in 1972 and, after a slow start, it  has become  a dominate theme in fundamental  physics. From time to time it has been suggested that supersymmetry is violated  by quantum corrections, that is, there is a supersymmetry anomaly. One of the problems with calculations in supersymmetric theories  is that there is no regulator which both preserves supersymmetry and gauge symmetry and this includes the most popular regulator, namely dimensional regularisation [1,2]. Dimensional regulation  preserves gauge symmetry and as a result  it played a key role in establishing the consistency of the standard model, however, it does not preserve supersymmetry. In the past the  claims that supersymmetry has an anomaly  have  been resolved by further explicit calculation of the offending Feynman diagrams which were causing the problem, see for example [3]. However the claims have also been disproved by establishing that the supersymmetry Ward identities really do hold if one took account of the fact that the regulator breaks supersymmetry [4,5]. Indeed in this way  it was shown that supersymmetry was not broken in $N=2$ and $N=4$ rigid supersymmetric models and in the Wess-Zumino model and super QED [4,5] for low numbers of loops. 
\par 
Supersymmetry anomalies do exist and they were first found [6] in the context of two dimensional models which also broke translation symmetry, that is, gravitational anomalies [7]. The supersymmetry and gravitational anomalies occurred in the same anomaly multiplet. A supersymmetry anomaly was found in four dimensions in reference [8]. However,  this appeared in theories  that had a gauge anomaly and as the calculation was performed in the Wess-Zumino gauge,  which is not a supersymmetric gauge choice,  the anomaly just reflected the presence of the gauge anomaly rather than a genuine supersymmetry anomaly. 
\par
Recently it has been proposed that when one couples rigid supersymmetric models to supergravity then one can find a supersymmetry anomaly and this occurs even for the free Wess-Zumino model coupled to supergravity [9-12]. Unlike the previously mentioned papers this work  included supergravity couplings. Indeed it was claimed that there was an anomaly in the free Wess-Zumino model in the presence of background supergravity fields [11]. A supersymmetric theory possesses spacetime translations and,  as  the presence of general relativity requires local translations,  it follows from the supersymmetry algebra  that  supersymmetry must also be  a local symmetry.  To have an anomaly in a local symmetry for which there are gauge particles, in this case the gravitini, renders the theory to be inconsistent. As a result if the claim of references [9-12] were to be true then the research on supersymmetry of the past fifty years would have been for nothing. It is therefore of some interest to see if the claims are true. Very recently certain  one loop current correlation functions of the  currents in the Wess-Zumino were calculated using  a Pauli-Villars regulator. It was found that there was no anomaly in the supersymmetry current [13] in these correlation functions. The results of references [9-12] were also further discussed in reference [14]. 
\par
In this paper we will consider the free Wess-Zumino model in the presence of a background supergravity background, in fact old minimal supergravity [15]. The corresponding path integral is 
$$
e^{iZ}=\int {\cal D}\varphi e^{iS^{WZ}+ j\cdot h} 
\eqno(1)$$
The action for the free Wess-Zumino model is given by 
$$
S^{WZ}=\int d^4 x \bigg \{-{1 \over 2}\left(\partial_\mu A \right)^2-{1 \over 2}\left(\partial_\mu B \right)^2-{1 \over 2}\bar \chi \partial \!\!\!/ \chi \bigg \}
\eqno(2)$$
and the integration ${\cal D}\varphi $ is over the fields in the  Wess-Zumino model, namely $A,B$ and $\chi_\alpha$. We do not need the auxiliary fields $F$ and $G$. We are using the conventions of reference [16] where a full explanation of many of the equations in this paper can also be found. The term $ j\cdot h$ represents the coupling of the fields of the Wess-Zumino model to the background supergravity fields and it is given by 
$$
j\cdot h\equiv i\int d^4 x (2h^{\mu\nu}\theta_{\mu\nu}+\bar \psi^{\mu\alpha }j_{\mu\alpha}- b^\mu j^{(5)}_\mu)
\eqno(3)$$
Here the supergravity fields $h^{\mu\nu}$, $ \psi^{\mu\alpha }$ and $b^\mu$ multiply the 
energy momentum tensor, the supersymmetry current and the axial current respectively.  We do not need the two auxiliary fields $M$ and $N$. For the free Wess-Zumino model the currents  are given by 
$$
\theta_{\mu \nu}={2 \over 3} \left(\partial_\mu A \partial_\nu A +\partial_\mu B \partial_\nu B \right) + {1 \over 4} \bar \chi \left(\gamma_\mu \partial_\nu + \gamma_\nu \partial_\mu \right) \chi
$$
$$-{1 \over 6}\eta_{\mu \nu} \left( \partial_\rho A \partial^\rho A + \partial_\rho B \partial^\rho B  \right)-{1 \over 3}A \partial_\mu \partial_\nu A-{1 \over 3} B \partial_\mu \partial_\nu B + {1 \over 3}\eta_{\mu\nu} \left( A \partial^2 A + B \partial^2 B \right)$$
$$j_\mu={1 \over 3}\partial \!\!\!/ \left(A-i\gamma_5 B \right) \gamma_\mu \chi +{2 \over 3} \partial_\mu \left( A+i\gamma_5 B \right) \chi-{2 \over 3}\left( A+i\gamma_5 B \right) \partial_\mu \chi +{2 \over 3} \left(A+i\gamma_5 B\right) \gamma_\mu \partial \!\!\!/ \chi $$
$$j_\mu^{(5)}={2 \over 3} \left( B \partial_\mu A-A \partial_\mu B +{i \over 4} \bar \chi \gamma_5 \gamma_\mu \chi \right)
\eqno(4)$$
It is straight forward to verify that they are conserved in the free Wess-Zumino model. These currents are known   to belong to a supermultiplet [17] and their supersymmetry transformations are given by $$
\delta j_\mu^{(5)}= i \bar \epsilon \gamma_5 j_\mu , \quad \delta \theta_{\mu \nu}=-{1 \over 4}\bar \epsilon \left( \gamma_{\mu \rho}\partial^{\rho}j_\nu+\gamma_{\nu \rho} \partial^{\rho}j_\mu \right), $$
$$
\delta j_\mu= \left(2\gamma^\nu \theta_{\mu \nu}-i \gamma_5 \partial  \!\!\!/  j_{\mu}^{(5)}-{i \over 2}\gamma_5\gamma_{\mu \kappa \lambda}\partial^{\kappa}j^{\lambda (5)} \right) \epsilon
\eqno(5)$$
The transformations of the old minimal supergravity fields at the linearised level can be found in that book [16]. 
\par
At the classical level,  the action and background coupling shown in equation (1) are supersymmetric in four dimensions. 
However, while this is true outside four dimensions for the free Wess-Zumino action it is not true for the field background supergravity coupling. Indeed the supersymmetry transformations of equation (4) are not true outside four dimensions as they require a Fierz reshuffle. One instead finds that under  a supersymmetry variation 
$$
\delta ( j\cdot h )\equiv G  = i \int d^4 x P^{\mu\alpha \beta} \partial_\mu \chi_\alpha \chi_\beta
\eqno(6)$$
plus a term that involves $\gamma^\mu j_\mu$ and one that involves $\partial^\mu j_\mu ^{(5)}$. We find that 
$$
P^{\mu\alpha \beta} = ( -{1 \over 3} \bar \epsilon^{\alpha} \left(\bar \psi^\lambda \gamma^\mu \gamma_\lambda \right)^\beta-{2 \over 3} \bar \epsilon^\beta \bar \psi^{\mu \alpha}-{2 \over 3} \bar \epsilon^\alpha \bar \psi^{\mu\beta})+ \big \{ \bar \epsilon \rightarrow \bar \epsilon i \gamma_5 , \bar \psi^\mu \rightarrow \bar \psi^\mu i \gamma_5 \big \}$$
$$-{1 \over 3}\bar \epsilon \gamma^\mu \psi^\nu \left(C\gamma_\nu \right)^{\alpha \beta}-{2 \over 3}\bar \epsilon \gamma^\nu \psi^\mu \left(C \gamma_\nu \right)^{\alpha \beta}-{1 \over 3} \bar \epsilon \gamma^\mu \gamma_5 \psi^\nu \left(C \gamma_\nu \gamma_5 \right)^{\alpha \beta}
\eqno(7)$$

\par
We now recall [4] the general procedure for calculating in a theory with a symmetry violating regulator  which we take to be dimensional regularisation but the procedure can be readily applied when other regulators are used.  In particular we take the spacetime dimension $D$ to be $D=4-2\epsilon$. We consider  a theory with fields $\varphi$ and an action $S$ whose variation is given by $\delta S= G$. The quantity $G$ is only non-zero due to the presence of the regulator and so it will vanish as the regulator is removed. In our case as $\epsilon \to 0$. The one particle irreducible effective action $\Gamma$ derived from the path integral with this action satisfies the Ward identity 
$$
\delta \varphi {\delta \Gamma\over \delta \varphi} = G\cdot \Gamma
 \eqno(8)$$
where the term on the right corresponds to  the insertion of  the operator $G$. To renormalise the theory we consider the quantities 
$$
\Gamma_{ren}= \Gamma-\Gamma_{div}-\Gamma ^{FLC} , \quad \quad (G\cdot \Gamma)_{ren} = G\cdot \Gamma- (G\cdot \Gamma)_{div}
 \eqno(9)$$
where $\Gamma_{div}$ and $(G\cdot \Gamma)_{div}$ are the divergent parts of  $\Gamma$ and $(G\cdot \Gamma)$ respectively 
and so they have a one over $\epsilon$ pole.  The term $\Gamma ^{FLC} $ allows for the possibility to add to the effective action a term which is finite and local in the fields. Clearly the coefficient of the pole does not depend on $ \epsilon$ and so we can take this to be in four dimensions and as such it  obeys the Ward identity. As $G$ vanishes as $\epsilon$ goes to zero the term $ (G\cdot \Gamma)_{ren} $ vanishes in this limit. However, $(G\cdot \Gamma)_{div}$ is of the form ${\epsilon \over \epsilon}$ and so this can be  finite and non-zero. We will find that in the limit as $\epsilon $ goes to zero the renormalised effective action obeys the Ward identity  
$$
\delta \varphi {\delta \Gamma_{ren}\over \delta \varphi} = 0
 \eqno(10)$$
 provided there do exist finite local counter terms in $\Gamma ^{FLC}$ which can be used to remove any parts of $(G\cdot \Gamma)_{div}$ which are not invariant under the symmetry. If this is not the case then the theory has an anomaly. 
 \par
 We now apply this general theory to supersymmetry and the free Wess-Zumino model in the background of supergravity fields reviewed above. In particular we will consider the terms in $(G\cdot \Gamma)_{div}$ which depend on the auxiliary field $b_\mu$. They also depend on the supersymmetry parameter $\epsilon^\alpha$ and the gravitino $\psi^\mu_\alpha$ as these appear in the expression for $G$ of equation (7). 
The  corresponding lowest order such Feynman diagram has two vertices, one contains the background gravitino and the two spinors $\chi_\alpha$ which form internal propagators. The other vertex contains the coupling $b^\mu j^{(5)}_\mu$ that appears in the original action of equation (1). The chiral current  $j^{(5)}_\mu$ given in  equation (4) contains the same two chiral spinors which also  form propagators. The resulting Feynman graph corresponds to the expression  
$$
{1\over 3} \int d^4p \int {d^4 k\over (2\pi)^4} {1\over k^2 (p+k)^2}Tr \{{ {k}}\!\!\!/ k_\mu CP^\mu (p) ( p \!\!\!/+ k\!\!\!/)\gamma_5\gamma_\nu \}b^\nu (-p)
\eqno(11)$$
where $C$ is the charge conjugation matrix. 
\par
The graph of equation (11) is relatively straight forward to evaluate using for example the formula in the classic Diagrammer [18] or the book [19]. One finds that it is equal to 
$$
-{2 \over 9} \int d^4 p {1 \over (4\pi)^{2}} {p^2 \over 12} \bigg \{-3 \bar \epsilon \gamma_5 b \!\!\!/ p \!\!\!/ \gamma \cdot \psi +2 \bar \epsilon \gamma_5 \gamma \cdot \psi b \cdot p + 2 \bar \epsilon \gamma_5 b \!\!\!/ p \cdot \psi - 2 \bar \epsilon \gamma_5 p \!\!\!/ b \cdot \psi \bigg \}
\eqno(12)$$
This can be removed by the  finite local counter term 
$$
\Gamma ^{FLC}={2 \over 9} \int d^4 p {1 \over (4\pi)^{2}} {p^2 \over 12} i \bigg \{ -\bar \psi_\mu p \!\!\!/ \psi^\mu + 2 \bar \psi \cdot p \gamma \cdot \psi + {5 \over 2} \bar \psi \cdot \gamma p \!\!\!/ \gamma \cdot \psi \bigg \} 
\eqno(13)$$
The two additional terms mentioned below equation (3) do not lead to any contribution to the effective action  and this corresponds to the fact that S-supersymmetry and the chiral current are conserved at least  for the processes being considered in this paper. 
As such there is no supersymmetry anomaly in this contribution to the effective action. 
\par
The calculation in this paper is only a first step in showing the consistency of supersymmetry in a supergravity  background using dimensional regularisation. It would be interesting to calculate the contribution containing the graviton.  Indeed as there are no gauge fields which require gauge fixing this calculation and the one in this paper  could be simultaneously  preformed in superspace using super Feynman rules. It may also be possible to completely evaluate the path integral of equation (1) as the contributions are only at one loop. 
\par
If one has gauge particles and one wants to calculate in components of familiar form then one must fix the Wess-Zumino gauge. However, this explicitly breaks supersymmetry. How to take account of this in the Ward Identities was explained in references [4] and [5]  and was successfully implemented in the context of the Wess-Zumino model, super QED and extended supersymmetry gauge theories although this was in the absence of background supergravity. It would be interesting to compare this approach to this problem with that of reference [20]. 

 In the paper [6] it was stated concerning in four dimensions that  "if the anomalies lie in a supermultiplet and the multiplet of currents is contained in a vector superfield then there are no supersymmetry anomalies. The proof of this theorem examines the most general vector multiplet that has a chiral current as its first component. One then identifies the most general possible energy-momentum tensor and demands that it be symmetric and conserved. The supervariation of these constraints imply that there exists a conserved supercurrent. " This statement would appear to transcend the theory being considered and it would mean that   there are no supersymmetry anomalies in four dimensions as there are no gravitational anomalies. It would be good to consider this in detail in the presence of background supergravity.   A review of possible anomaly multiplets can be found in chapter twenty of reference [16] although this was before the discovery of supersymmetry anomalies.


\medskip
{\bf {Acknowledgments}}
\medskip
I wish to thank  the SFTC for support from Consolidated grants number ST/T000759/1  and ST/P000258/1.

\medskip
{\bf {References }}
\medskip
\item{[1]} J. Ashmore, {\it A Method of Gauge Invariant Regularization}, Nuovo Cimento Lett. 4 (1972) 289.
\item{[2]} C. Bollini and J. Giambiagi, {\it Dimensional Renormalization: The Number of Dimensions as a Regularizing Parameter}, Nuovo Cimento 12B (1972) 20;
\item{[3]} I. Jack, {\it Two Loop Beta Functions for Supersymmetric Gauge Theories}, Phys. Lett. 147B (1984) 405
\item{[4]} P.S. Howe, A.J. Parkes and P.C. West, {\it Dimensional Regularisation and Supersymmetry}, Phys. Lett. 147B (1984) 409.
\item{[5]} P.S. Howe, A.J. Parkes and P.C. West, {\it On Supersymmetry Anomalies}, Phys. Lett. 150B (1985) 149. 
\item{[6]} P. Howe and P. West, {\it Gravitational Anomalies in Supersymmetric Theories},  Phys. Lett. {\bf156B} 335 (1985).
\item{[7]} L. Alvarez-Gaume and E. Witten, {\it Gravitational anomalies}, Nucl. Phy~ B234 (1984) 269.
\item{[8]} H. Itoyama, V. P. Nair, and H.-c. Ren, Nucl. Phys.{\it Supersymmetry Anomalies and some aspects of renormalisation},  B262, 317 (1985).
\item{[9]}  I. Papadimitriou, {\it Supercurrent anomalies in 4d SCFTs}, JHEP07(2017) 038, ArXiv:1703.04299.
\item{[10]}  I. Papadimitriou, {\it Supersymmetry anomalies in new minimal supergravity}, JHEP09(2019)039 [1904.00347].
\item{[11]}  G. Katsianis, I. Papadimitriou, K. Skenderis and M. Taylor, {\it Anomalous Supersymmetry}, Phys. Rev. Lett.122(2019) 231602,  ArXiv:1902.06715.
\item{[12]}  I. Papadimitriou, {\it Supersymmetry anomalies inN= 1conformal supergravity}, JHEP04(2019) 040. ArXiv:1902.06717.
\item{[13]} A. Bzowski, G.  Festuccia and V. Prochazka. {\it Consistency of supersymmetric Õt Hooft anomalies}, ArXiv:2011.09978
\item{[14]} G. Katsianis, I. Papadimitriou, K. Skenderis and M. Taylor, {\it Supersymmetry anomaly in the superconformalWess-Zumino model} 
\item{[15]} K.S. Stelle and P. West. {\it Minimal Auxiliary Fields for Supergravity}, Phys. Lett {\bf74B}, 330 (1978); 
S. Ferrara and P. van Nieuwenhuizen, {\it The Auxiliary Fields of Supergravity}, Phys. Lett. {\bf B74}, 333 (1978). 
\item{[16]} P. West,  {\it Introduction to Supersymmetry and Supergravity}, World Scientific Publishing 1986, second edition 1990. 
\item{[17]}  S. Ferrara and B. Zumino, {\it Transformation Properties of the Supercurrent}, Nucl. Phys. B87(1975) 207.
\item{[18]} G. 't Hooft and M. Veltman, {\it Diagrammer}, CERN preprint 73-9, 3 September 1973. 
\item{[19]} P. Ramond, {\it Field Theory: A Modern Primer}, Benjamin/Cummings 1981. 
\item{[20]}  S.M. Kuzenko, A. Schwimmer and S. Theisen, {\it Comments on Anomalies in SupersymmetricTheories}, J. Phys. A53 (2020) 064003,  ArXiv:1909.07084.

\end